\documentclass[11pt]{article}
\usepackage[
top    = 2.50cm,
bottom = 2.50cm,
left   = 2.50cm,
right  = 2.50cm]{geometry}

\usepackage[pdftex]{graphicx}
\usepackage{tabularx}
\usepackage{latexsym}
\usepackage{amsmath}
\usepackage{amssymb}
\usepackage{cite}

\usepackage{amsfonts}
\newcommand{\abs}[1]{\lvert #1 \rvert}
\newcommand{\avg}[1]{\left\langle {#1} \right\rangle}

\newcommand{\point}{\hspace{0.3cm}.}
\newcommand{\psik}{\hspace{0.3cm},}

\newcommand{\heading}[1]{\begin{center} \Large {#1} \end{center}}
\usepackage{multirow}
\usepackage{pdfpages}

\def\tilde{\widetilde}

\makeatletter
\newcommand*\rel@kern[1]{\kern#1\dimexpr\macc@kerna}
\newcommand*\widebar[1]{%
  \begingroup
  \def\mathaccent##1##2{%
    \rel@kern{0.8}%
    \overline{\rel@kern{-0.8}\macc@nucleus\rel@kern{0.2}}%
    \rel@kern{-0.2}%
  }%
  \macc@depth\@ne
  \let\math@bgroup\@empty \let\math@egroup\macc@set@skewchar
  \mathsurround\z@ \frozen@everymath{\mathgroup\macc@group\relax}%
  \macc@set@skewchar\relax
  \let\mathaccentV\macc@nested@a
  \macc@nested@a\relax111{#1}%
  \endgroup
}

\begin{document}

\begin{titlepage}
  \vspace*{\fill}

\heading{\bf New Phases in O'Raifeartaigh-like Models and $R$-symmetry Breaking}

\vskip 1.4cm

\centerline{\it Talya Vaknin }
\bigskip
\centerline{ Weizmann Institute of Science, Rehovot
76100, Israel}

\vskip 4pt

\vskip 1.5cm

\begin{abstract}

We study $\mathcal{N}=1$ models in which supersymmetry and $R$-symmetry are broken spontaneously. We find that even the simplest well-known O'Raifeartaigh-like models exhibit previously unnoticed phases and dynamics. D-terms play an important role in the dynamics, creating a runaway behavior at tree-level which is stabilized radiatively at a large scale. This leads to a dynamically-generated hierarchy of scales.  Additionally, this mechanism generically leads to a spontaneously broken  $R$-symmetry without  fine tuning the couplings of the theory. These are arguably the simplest known models that break the $R$-symmetry spontaneously.

\end{abstract}

  \vspace*{\fill}

\end{titlepage}

\pagebreak

\tableofcontents

\section {Introduction}

 There are many reasons to believe that supersymmetry is, ultimately, a symmetry of the physical laws. The energy scale at which supersymmetry is broken is unknown and has been a topic of extensive research in the last few decades. There are several types of ``toy models,'' or ``classes'' of supersymmetry breaking models. These different models may sometimes provide the basis for constructing realistic phenomenological models.

In this note we are primarily interested in a simple class of toy models of supersymmetry breaking, namely those in which non-perturbative corrections play no significant role. While in and by themselves they are not particularly interesting, they often arise as the low energy effective theory of strongly coupled field theories (see e.g.~\cite{ISSmeta2006, Izawa, IT}). Hence, it is crucial to understand the possible dynamics of such simple perturbative models.  As we will show in this work, somewhat surprisingly,  there are previously unnoticed important facts about such perturbative models.

The simplest prototype of a calculable, perturbative model that breaks supersymmetry was discovered by O'Raifeartaigh (O'R) more than three decades ago~\cite{O'R:1975} . Many authors studied generalizations of this model, including ones that are useful for phenomenology. A common feature of such models~\cite{Ray,Komargodski_Shih} is the existence (at tree-level) of infinite degeneracy of SUSY-breaking vacua. The fate of such flat directions is decided by radiative corrections. Such flat SUSY-breaking directions that exist at tree-level are called {\it pseudomoduli}. One interesting application of these flat directions was discussed in \cite{Witten:mass}, where it was demonstrated that the vacuum of the theory can reside at a scale much larger than all the fundamental scales of the O'R model, thereby potentially explaining the remarkable hierarchy between the GUT scale and the electroweak scale.

Spontaneous SUSY breaking implies the existence of an $R$-symmetry \cite{Nelson}, which, if it were unbroken, forbids gaugino mass terms. According to experimental data, gauginos should be massive therefore $R$-symmetry must be broken predominantly spontaneously. This is another constraint that needs to be borne in mind if we are after realistic models.

We consider the simplest low-energy models in which $R$-symmetry and supersymmetry break spontaneously. Some examples for $R$-symmetry breaking at tree level are \cite{Sun} and \cite{carpenter}, a more  general analysis of these models is given by \cite{Komargodski_Shih}. $R$-symmetry can also be broken at one loop \cite{Shih_R}, \cite{ISSmeta} or higher loops \cite{Intriligator:2008fe,Giveon_Komargodski,Amariti_2loops}. Most of the existing literature considered spontaneous SUSY breaking via F-terms alone, or where D-terms did not play a major role. This leads to anomalously small gaugino masses \cite{Komargodski_Shih}.

We may ask, can D-terms can play a role in SUSY breaking at all? Can they change the dynamics of the model?
There is a well-known theorem that states that D-terms can be set to zero as long as the F-terms have a solution.\footnote{See ref. \cite{Wess}. This is a result of a more general theorem, which states that the space of D-flat directions is isomorphic to the space of holomorphic gauge invariants \cite{Buccella,Procesi,Gatto,Luty}}
 A generalization of this result appeared in~\cite{Dumitrescu}. This gives the impression that they might not have an important role in SUSY breaking.
On the other hand there are examples in which D-terms can become important and comparable to the F-terms. See for example \nolinebreak ~\cite{Dumitrescu, Rattazzi, Romanino, Shacham, Azeyanagi}.

Here we will study models where D-terms lead to dynamics remarkably different from known examples. We describe simple models where the breaking of $R$-symmetry is achieved effortlessly at one loop, and the breaking is parametrically large. This is phenomenologically desirable and different from some previous one loop mechanisms for breaking $R$-symmetry (which happen to be somewhat tuned) \cite{ISSmeta}. Moreover, the fields in our model have $R$-charge $0$ or $2$; for models with this $R$-charge assignment there are many known UV dynamical completions. For theories with other $R$-charge assignments see the one loop mechanism of \cite{Shih_R}. In addition, as mentioned above, D-terms play an important role in our analysis, providing more examples of the possible role of D-terms in SUSY breaking.

 Our study is based on an observation by~\cite{Matos}, where it was argued that gauging some global symmetries of a theory generically leads to a runaway direction at tree level. We find that at one-loop there is a minimum along this runaway which breaks supersymmetry and $R$-symmetry. The simplest realization of a theory where this takes place is identical to the original (vector-like version) of the O'R model. This model has been studied many times before, indeed it is the simplest model of SUSY breaking. However, it appears that such a fundamental feature in its phase diagram was overlooked.
Phenomenologically, an application of our study could be to utilize this new minimum in order to explain some mild hierarchy problems (such as in split supersymmetry with the sfermions at $10^4$ TeV, a model that is receiving nowadays some interest due to the Higgs-like particle at 125 GeV \cite{Romanino}) in the spirit of \cite{Witten:mass}.

The simple example we analyze in detail in this paper admits various generalizations. In fact, one can argue that the existence of such phases is rather generic.

 The outline of this paper is as follows. We begin with a review of the O'R model (before gauging) and remind how it breaks SUSY but not $R$-symmetry. Then we gauge the theory and show that it has a runaway behavior that is stabilized by one-loop corrections. This leads to a spontaneous breaking of both SUSY and $R$-symmetry. We finish by using gauge invariant variables in order to calculate the masses of the fields in the theory.

     \section{O'Raifeartaigh Model (no Gauging)}

   In this section we will examine the familiar O'R model, in which there is no gauge dynamics. We will  review that in this case SUSY is broken but $R$-symmetry is preserved by the vacuum.
   The superpotential is given by:
\begin{equation}
   W=hX(\phi\tilde{\chi}-\mu^{2})+m_{1}\phi\tilde{\phi}+m_{2}\chi\tilde{\chi} \psik
\end{equation}
 with canonical K\"ahler potential.\\
  This model admits a global $U(1)$ symmetry under which two chiral superfields ($\phi$, $\chi$) are positively charged, two chiral fields are negatively charged ($\tilde\phi$,$\tilde\chi$), and $X$ is  singlet. Additionally, there is an $R$-symmetry under which $X,\tilde\phi,\chi$ carry charge $2$, while the rest are neutral.
 We will not discuss the natural $SO(N)$ generalization of this model here.
     The scalar potential is
   \begin{equation}
   \begin{aligned}
   \label{eqn:vf}V_{F}&=\abs{F_X}^2+\abs{F_\chi}^2+\abs{F_{\tilde\phi}}^2+\abs{F_\phi}^2+\abs{F_{\tilde\chi}}^2\\
   &=h^{2}\abs{\phi\tilde{\chi}-\mu^{2}}^{2}+m_{2}^{2}\abs{\tilde{\chi}}^{2}+m_{1}^{2}\abs{\phi}^{2}+\abs{hX\tilde{\chi}+m_{1}\tilde{\phi}}^{2}+\abs{hX\phi+m_{2}\chi}^{2}\point
   \end{aligned}
   \end{equation}
   SUSY is broken at tree level since we cannot set all terms to zero simultaneously: when setting $F_{\tilde\phi}=0$ we automatically get $F_X\neq 0$.

  The potential does have a supersymmetry-breaking minimum, we can find it in the usual way.
   The last two terms of $V_F$ can be set to zero at no energy cost, from them we get the relations:
    \begin{gather}
   \label{eqn:phi}\tilde{\phi}=-\frac{hX}{m_1}\tilde{\chi}\psik\\
   \label{eqn:chi}\chi=-\frac{hX}{m_2}\phi\point
   \end{gather}
   Therefore we are left to find the values of $\phi$, $\tilde{\chi}$ and $X$ which minimize $\abs{F_X}^2+\abs{F_\chi}^2+\abs{F_{\tilde\phi}}^2$.
   There are two phases in this model. In the first one, $\phi=0$ and $\tilde{\chi}=0$, while $X$ is undetermined. In this case $V_{F}=h^{2}\mu^{4}$ and it is the absolute minimum in the regime $\mu^{2}<\frac{m_{1}m_{2}}{h^{2}}$. Since all the charged fields have zero vevs, the $U(1)$ symmetry is unbroken, therefore we will call this phase the {\it unbroken phase}. The $R$-symmetry is unbroken at $X=0$ but it is broken elsewhere.

   In the second phase, it can be shown that the solutions are real fields and they get the values
   \begin{equation}
   \begin{split}
   \label{eqn:vevs1}\phi&=-\frac{m_{2}y}{h}\\
   \tilde{\chi}&=-\frac{m_{1}y}{h}\psik
   \end{split}
   \end{equation}
   with $y=\frac{\sqrt{h^2\mu^2-m_1m_2}}{\sqrt{m_1m_2}}$.
   From (\ref{eqn:phi}) and (\ref{eqn:chi}) we immediately get $\chi=\tilde{\phi}=yX$
   and $X$ stays undetermined. In this second case, the $U(1)$ is broken while the $R$-symmetry is unbroken at $X=0$ but broken elsewhere. The vacuum energy is $V_{F}=2\mu^{2}m_{1}m_{2}-\frac{m_{1}^{2}m_{2}^{2}}{h^{2}}$, it is the global minimum in the regime $\mu^{2}>\frac{m_{1}m_{2}}{h^{2}}$. We will refer to this phase as the {\it broken phase}.

   To summarize, the global minima of the potential are
   \begin{equation}
   \label{eqn:vzero}V_{F}=\left\{ \begin{array}{lll}
   h^{2}\mu^{4} & \psik & \mu^{2}<\frac{m_{1}m_{2}}{h^{2}}\\
   2\mu^{2}m_{1}m_{2}-\frac{m_{1}^{2}m_{2}^{2}}{h^{2}} & \psik & \mu^{2}>\frac{m_{1}m_{2}}{h^{2}}
   \end{array}\right. \point
   \end{equation}
   The transition between these two phases is a second order phase transition i.e. the energy density varies smoothly, but not its first derivatives.

   We will now examine the dynamics of the undetermined pseudomodulus $X$ when considering one-loop corrections \nolinebreak \cite{O'R:1975}. We will review the known result that the degeneracy of  the vacuum is lifted in such a way that the $R$-symmetry is unbroken in both phases.

The one-loop effective potential is given by \cite{Coleman:1973}
\begin{equation}
\begin{aligned}
\label{eqn:col_wein}V_{eff}^{(1)}&=\frac{1}{64\pi^2}STr\Bigl ( \mathcal M^4 log\frac{\mathcal M^2}{m_0^2} \Bigr )\\
&=\frac{1}{64\pi^2}\Bigl[Tr \Bigl(m_B^4 log\frac{m_B^2}{m_0^2}\Bigr )-Tr\Bigl (m_F^4 log\frac{m_F^2}{m_0^2}\Bigr)\Bigr]\psik
\end{aligned}
\end{equation}
where $m_0$ is the SUSY breaking scale. In the limit $X\approx 0$ it takes the form $V_{eff}=const+m_X^2 X^2+O(X^3)$ in both regimes of (\ref{eqn:vzero}). $m_X$ is just a constant depending on the masses of the model and is different in the two regimes.
Since $m_X^2$ is positive, the pseudomodulus has a minimum at $\avg{X}=0$.

In the limit $X\gg m_0$ we can use the result given in \cite{Intriligator:2008fe} where the full expression (\ref{eqn:col_wein}) is approximated by the contribution only to the effective K\"ahelr potential:
\begin{equation}
\label{eqn:Veff} V_{eff}(X)\approx const.+2V_0\gamma log\frac{\abs{X}}{m_0}\psik
\end{equation}
$V_0$ is the tree level vacuum energy, given by (\ref{eqn:vzero}) and $\gamma$ is the anomalous dimension, which is positive.
We conclude that for $X\rightarrow\infty$ the one loop correction is proportional to $log(\abs{X})$, and so is an increasing function of $X$ (for large enough $X$).

Finally, using the full expression from (\ref{eqn:col_wein}), it can be shown that the effective potential is monotonic between these two limits. Therefore, we conclude that $\avg{X}=0$ is the global minimum of the potential and in both phases, $R$-symmetry is unbroken.

 \section{Gauging the $U(1)$, Breaking $R$-Symmetry Spontaneously}

 When gauging the $U(1)$ symmetry, $R$-symmetry can be broken in various ways in this model. We begin with a general analysis which shows that there is a runaway behavior at tree level in both phases of the model. This runaway happens to be a general phenomena of gauged theories in which the F-flatness conditions are not satisfied \cite{Japan2011, Matos}. Then we review the results of Matos \cite{Matos} who showed a runaway at the broken phase. Furthermore, we give an example of a runaway in the unbroken phase. Then, we will show that all these runaways are stabilized at one loop and we get a hierarchically large breaking of $R$-symmetry enhanced by a loop factor. Finally, we'll turn to examine the behavior of the potential at the origin and show that due to the gauging, it can be smoothly connected to the runaway that we found.
\\

After gauging the $U(1)$ symmetry in the model, the full potential is $V_F+V_D$ with $V_F$ given by \nolinebreak (\ref{eqn:vf}) and
\begin{equation}
\label{eqn:VD} V_{D}=\frac{g^{2}}{2}(\xi+\abs{\phi}^{2}-\abs{\tilde{\phi}}^{2}+\abs{\chi}^{2}-\abs{\tilde{\chi}}^{2})^2\psik
\end{equation}
where $g$ is the gauge coupling and $\xi$ is the Fayet-Iliopoulos term.

\subsection{Runaway to $V_D=0$}
\label{sec:runaway}
\begin{figure}[h]
   \centering
   \includegraphics[width=0.5\textwidth]{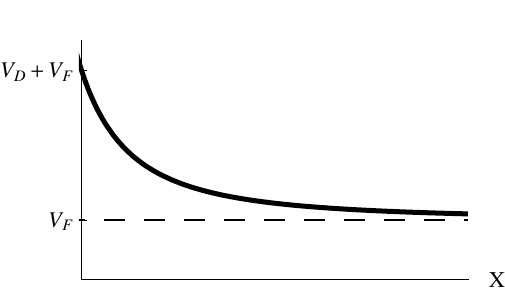}
   \caption{The potential at large $X$ admits a runaway behavior: When $X\rightarrow\infty$ the potential approaches the value of the SUSY breaking vacuum of the ungauged theory, $V_F$. }
   \label{plot:V}
   \end{figure}

We shall denote the vevs of the fields $\phi$ and $\tilde\chi$ by $\phi_0$ and $\tilde\chi_0$ respectively so that we can carry out the analysis for both phases simultaneously. We can deform the vevs of the fields (\ref{eqn:vevs1}) in the following way:
\begin{equation}
\begin{gathered}
\avg{\phi}=\phi_0+\epsilon_1\\
\avg{\tilde\phi}=-\frac{hX}{m_1}\avg{\tilde\chi}+\eta_2\\
\avg{\chi}=-\frac{hX}{m_2}\avg{\phi}+\eta_1\\
\avg{\tilde\chi}=\tilde\chi_0+\epsilon_2\point
\end{gathered}
\end{equation}
By doing this deformation the scalar potential is:
\begin{align}
 \label{eqn:runaway}V=V_F^0+\mathcal O(\epsilon_i,\eta_j)+\frac{g^2}{2}\Bigl[\sqrt{V_D^0}-h\Bigl(\frac{X(\phi_0+\epsilon_1)\eta_1^*}{m_2}-\frac{X(\tilde\chi_0+\epsilon_2)\eta_2^*}{m_1}+c.c\Bigr)\\
 +h^2\abs{X}^2\Bigl(\frac{\abs{\epsilon_1}^2}{m_2^2}-\frac{\abs{\epsilon_2}^2}{m_1^2}+(\frac{\epsilon_1\phi_0^*}{m_2^2}-\frac{\epsilon_2\tilde\chi_0^*}{m_1^2}+c.c)\Bigr)+\mathcal O(\epsilon_i,\eta_j)\Bigr]^2\point
\end{align}
Where $i,j=1,2$ and $V_F^0$ and $(\frac{g^2}{2})^{-1}V_D^0$ are the scalar potentials with no deformations.
We see that in both phases we can choose $\epsilon_i$ and $\eta_j$ such that for very large $\abs{X}$, the potential exhibits a runaway behavior to $V_D^0=0$ (figure \ref{plot:V}). This behavior was presented in \cite{Matos} for the broken phase choosing $\eta_1=-\eta_2=\eta$, $\epsilon_i=0$ and $X=\frac{y^2(m_1^2-m_2^2)-h^2\xi}{4h^2y\eta}$.
For a runaway in the unbroken phase we can choose, for example $\epsilon_1=\epsilon_2=\epsilon$, $\eta_j=0$ and $\abs{X}^2=\frac{\xi}{h^2\epsilon^2}\bigl(\frac{1}{m_1^2}-\frac{1}{m_2^2}\bigr)^{-1}$.

Notice that in the broken phase, the parameter $\xi$ and the difference $m_1-m_2$ play the same role in the dynamics of the model, therefore we can set either one of them (but not both) to zero and still have a runaway. However, to get a runaway in the unbroken phase, we must introduce a FI term: If $\xi=0$ in the unbroken phase, $V_D=0$ so the D-terms don't play a role in the dynamics of this model; precisely, there is no runaway.

\subsection{One loop corrections}

We examine how the one loop corrections affect these runaway directions. At first, we will calculate the corrections at leading order in $g$. We assume $X\gg m_0$ therefore we can use the approximated one loop contribution (\ref{eqn:Veff}), and insert $V_0$ from (\ref{eqn:vzero}) and the anomalous dimension which is given by
\begin{equation}
\label{eqn:anom_dim}\gamma=\frac{\abs{F_X}^2\gamma_X+\abs{F_{\tilde \phi}}^2\gamma_{\tilde \phi}+\abs{F_\chi}^2\gamma_\chi+\abs{F_\phi}^2\gamma_\phi+\abs{F_{\tilde\chi}}^2\gamma_{\tilde\chi}}{\abs{F_X}^2+\abs{F_{\tilde \phi}}^2+\abs{F_\chi}^2+\abs{F_\phi}^2+\abs{F_{\tilde\chi}}^2}\psik
\end{equation}
  estimated at the vevs along the runaway (\ref{eqn:vevs1}). This results in $\gamma=\frac{h^2}{32\pi^2}\frac{m_1m_2}{2\mu^2h^2-m_1m_2}$ for the broken phase and $\gamma=\frac{h^2}{32\pi^2}$ for the unbroken phase.
The effective potential is then approximated by
\begin{equation}
\Delta V\approx\left\{ \begin{array}{lll}
   \frac{h^{4}\mu^{4}}{16\pi^2}log\frac{\abs{X}}{m_0} & \psik & \mu^{2}<\frac{m_{1}m_{2}}{h^{2}}\\
   \frac{m_1^2m_2^2}{16\pi^2}log\frac{\abs{X}}{m_0} & \psik & \mu^{2}>\frac{m_{1}m_{2}}{h^{2}}
   \end{array}\right. \point
\end{equation}
If we follow Matos's choice of deformation for the broken phase, then the full potential takes the form
\begin{equation}
V\approx 2m_1m_2\mu^2-\frac{m_1^2m_2^2}{h^2}+\frac{y^2}{4h^2X^2}(m_1^2+m_2^2)(m_1^2-m_2^2)^2+\frac{m_1^2m_2^2}{16\pi^2}log\frac{\abs{X}}{m_0}\point
\end{equation}
This potential has a minimum at
\begin{equation}
\label{eqn:minima} h^2X^2\approx\frac{2\pi^2}{ h^2}\frac{(m_1^2-m_2^2)^2(m_1^2+m_2^2)(h^2\mu^2-m_1m_2)}{m_1^3m_2^3}\point
\end{equation}

We can do a similar analysis for the unbroken phase, going back to section (\ref{sec:runaway}), and using the example given there, we get
\begin{equation}
V\approx h^2\mu^4+\frac{\xi}{h^2X^2}(m_1^2+m_2^2-2h^2\mu^2)(\frac{1}{m_1^2}-\frac{1}{m_2^2})^{-1}+\frac{\xi^2}{h^2X^4}(\frac{1}{m_1^2}-\frac{1}{m_2^2})^{-2}+\frac{h^{4}\mu^{4}}{16\pi^2}log\frac{\abs{X}}{m_0} \point
\end{equation}
In this case, the minimum is balanced at
\begin{equation}
h^2X^2\approx \frac{8\pi\xi}{h^4\mu^4}(\frac{1}{m_2^2}-\frac{1}{m_1^2})^{-1}\Bigl(2\pi(2h^2\mu^2-m_1^2-m_2^2)+\sqrt{h^6\mu^4+4\pi^2(2h^2\mu^2-m_1^2-m_2^2)^2}\Bigr)
\end{equation}

We see that in both phases the runaway direction is stabilized at $\avg{X}\neq 0$, this global minimum breaks $R$-symmetry as well as supersymmetry. Moreover, a large energy scale is generated dynamically, it is enhanced by a loop factor compared to the scales that appear at tree level. This ensures that the approximation used to obtain this minimum is self consistent.

The next order in $g$ enters into the effective potential through the anomalous dimension.
Going back to (\ref{eqn:anom_dim}), we get anomalous dimensions for the $\tilde\phi$ and $\chi$ fields as well, with values $\gamma_{\tilde\phi}=\gamma_\chi=-\frac{g^2}{8\pi^2}$.
In the unbroken phase, there is no correction since $F_{\tilde\phi}=F_{\chi}=0$. In the broken phase, the anomalous dimension is corrected to
\begin{equation}
\gamma=\frac{1}{4\pi^2}\frac{\frac{1}{8}m_1^2m_2^2-\frac{g^2}{h^2}m_1m_2(h^2\mu^2-m_1m_2)}{2\mu^2m_1m_2-\frac{m_1^2m_2^2}{h^2}}\point
\end{equation}
Therefore, the one-loop correction to the effective potential is
\begin{equation}
\Delta V=\Bigl(\frac{m_1^2m_2^2}{16\pi^2}-\frac{g^2}{h^2}\frac{m_1m_2}{2\pi^2}(h^2\mu^2-m_1m_2)\Bigr)log\frac{\abs{X}}{m_0}\point
\end{equation}
This result is compatible with the result in \cite{Dine_Mason} when considering two different masses. Only when the $log$ has a positive coefficient does the runaway stabilize and get a minimum. We see that at this order in $g$ there is a minimum only as long as $(g^2/h^2)(h^2\mu^2-m_1m_2)<1/8$.

To conclude, the theory has a tree-level runaway which is stabilized by one-loop effects. In the broken phase, $g$ must be smaller than a certain combination of the mass scales of the theory in order to get a stable minimum along the runaway.

\subsection{Potential at the origin}

We turn to examine the behavior of the potential at the origin and see how it can be embedded into a coherent picture along with the runaway. Recall that before gauging, there is a stable minimum at $X=0$.

\begin{figure}[h]

   \centering
   \includegraphics[width=0.5\textwidth]{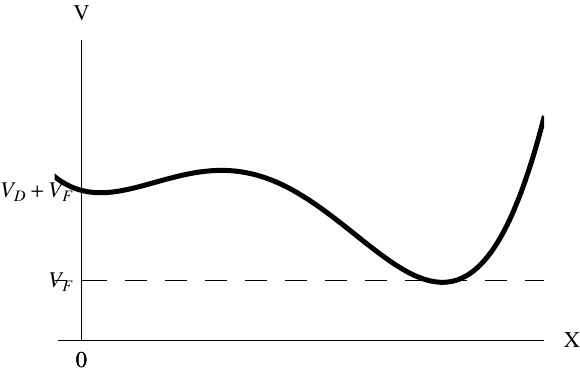}
   \caption{The potential in the unbroken phase: One loop corrections give rise to a minimum along the runaway, at large $X$. This is the absolute minimum of the potential, while the minimum at the origin becomes meta stable. }
   \label{plot:V2}
   \end{figure}

In the unbroken phase, the gauging does not affect the minimum at the origin since $V_D$ is just a constant. Therefore, the minimum at the origin is equal to $V_F+V_D$ while on the runaway $V_D\longrightarrow 0$. The minimum at the origin is meta stable while the R- breaking minimum is the absolute minimum of the theory (figure \ref{plot:V2}).
Notice that this meta stable minimum is long lived, having the distance between the minimums enhanced by a one loop factor. Unlike other theories, where the meta stable minimum is made long lived by introducing a small scale (for example \cite{ISSmeta2006}), in this model it arises dynamically.

In the broken phase, in order to examine the behavior at the origin when considering gauging, we will consider small fluctuations of the vevs (\ref{eqn:vevs1}) at $X=0$

\begin{equation}
\begin{gathered}
\phi=\phi^0+\delta\phi\\
\tilde\phi=\delta\tilde\phi\\
\chi=\delta\chi\\
\tilde\chi=\tilde\chi^0+\delta\tilde\chi\point
\end{gathered}
\end{equation}
Where $\phi^0$ and $\tilde\chi^0$ are the vevs we found (\ref{eqn:phi}) and (\ref{eqn:chi}).
When inserting these into the scalar potential, we get tadpole contributions from the $R$-uncharged fields $\phi$ and $\tilde\chi$:
\begin{equation}
V_D=const+2g^2\Bigl((\phi^0)^2-(\tilde\chi^0)^2+\chi\Bigr)
(\phi^0\delta\phi+\tilde\chi^0\delta\tilde\chi)+...\point
\end{equation}
Where the ellipsis stand for higher terms in the fields' fluctuations.
Due to this tadpole contribution, the minimum is shifted away from its original values. We will now show that at least one of the $R$-charged fields becomes tachyonic at this point: We begin by calculating the shifted vevs of $\phi$ and $\tilde\chi$ by solving the equations $\frac{\partial V}{\partial\phi}=0$ and $\frac{\partial V}{\partial \tilde\chi}=0$ at the origin, i.e. at $X=\tilde\phi=\chi=0$. We get
\begin{equation}
\begin{aligned}
\label{eqn:vevsg4}\phi=& -\frac{m_2y}{h}+g^2\frac{(y^2(m_1^2-m_2^2)+h^2\xi)(m_2^2(y^2-1)+m_1^2(y^2+1))}{4h^3m_1m_2^2y}\\ &+\frac{g^4}{32h^5m_1^4m_2^3y^3}(y^2(m_1^2-m_2^2)-h^2\xi)\\&
\Bigl(y^2\bigl(m_1^6(7y^4-3)+m_1^4m_2^2(7y^4-12y^2+9)+9m_1^2m_2^4(y^4-1)+3m_2^6(3y^4+4y^2+1)\bigr)\\&
-h^2\xi\bigl(m_1^4(3y^4+1)-2m_1^2m_2^2(y^2+1)^2-m_2^4(5y^4+4y^2-1\bigr)\Bigr)
+\mathcal O(g^6)\\
\tilde\chi=&(m_1\leftrightarrow m_2)\point
\end{aligned}
\end{equation}
Then we insert these into the quadratic part of the scalar potential for the $R$-charged fields:
\begin{equation}
\begin{pmatrix} X^*&\tilde\phi^*&\chi^*\end{pmatrix}m^2
\begin{pmatrix}X\\ \tilde\phi\\ \chi\\ \end{pmatrix}\psik
 \end{equation}
with
\begin{equation}
m^2=
\begin{pmatrix}
h^2(\abs{\phi}^2+\abs{\tilde\chi}^2)&hm_1\tilde\chi^*&hm_2\phi^*\\
hm_1\tilde\chi&m_1^2-g^2(\abs{\phi}^2-\abs{\tilde\chi}^2+\xi)&0\\
hm_2\phi&0&m_2^2+g^2(\abs{\phi}^2-\abs{\tilde\chi}^2+\xi)
\end{pmatrix}\point
\end{equation}
and calculate the determinant
\begin{equation}
det(m^2)=\frac{-2g^2y^2(m_1^2+m_2^2)}{h^4}(m_1^2y^2-m_2^2y^2-h^2\xi)^2\point
\end{equation}
Unless $y=0$, the determinant is negative therefore there is a negative eigenvalue. In other words, upon gauging, the origin $X=0$ is no longer a stable minimum of the theory.

We conclude that in the broken phase, the behavior of the potential near the origin depends on the ratio between the gauge coupling $g$ and the one-loop contribution: If one-loop effects are larger than $g$, the pseudomodulus is lifted and we get a stable minimum at $X=0$ but if $g$ is larger than one-loop effects, $X=0$ is no longer a stable minimum. Hence we expect to have a critical $g$ for which there is a phase transition between these regimes.
It seems that there are no local minima although we have not proved this.
Combing this with the restrictions on $g$ from the one loop calculations we conclude that in the broken phase, the theory has a runaway which is stabilized only for certain gauge couplings: $g$ must be larger than one loop effects in order to get a runaway behavior. In addition, it must be smaller than a certain combination of the mass scales of the theory in order to get a stable minimum along the runaway. This regime in parameter space is large and contains phenomenologically familiar values for $g$.

\section{Field Masses}

Along the runaway direction $V_D$ is small, therefore we can switch to gauge invariant variables, for convenience we set $\xi=0$.
In this approximation we can easily calculate the masses of the fields.

We will switch to these holomorphic gauge invariant binomials: $M=\phi\tilde\chi, L=\sqrt{\tilde\phi\chi}, P=\phi\tilde\phi, R=\chi \tilde\chi$, these satisfy a simple relation $ML^2-PR=0$.\\
The K\"ahler  potential and superpotential along the D-flat direction have the following form:
\begin{align}
K&=X^\dagger X+2\sqrt{M^\dagger M+(L^\dagger L)^2+P^\dagger P+ R^\dagger R}\psik\\
W&=hX(M-\mu^2)+m_1P+m_2R\point
\end{align}
The constraint can be used to solve for $M$ ($M=PR/L^2$):
\begin{align}
K&=X^\dagger X+2\sqrt{(L^\dagger L)^2+P^\dagger P+R^\dagger R+ \frac{(PR)^\dagger(PR)}{(L^\dagger L)^2}}\psik\\
W&=hX(\frac{RP}{L^2}-\mu^2)+m_1P+m_2R\point
\end{align}

We can now take some limits in order to evaluate the fields masses at the minimum that we found along the runaway (\ref{eqn:minima}).
For large $X$ we can integrate out $P$ and $R$ and approximate the equations of motion by: \\
\begin{equation}
   \begin{aligned}
\label{eom_massivefields} 0&=\frac{\partial W}{\partial R}=\frac{hXP}{L^2}+m_2\psik\\
0&=\frac{\partial W}{\partial P}=\frac{hXR}{L^2}+m_1\point
\end{aligned}
\end{equation}
This results in:
\begin{align}
K&=X^\dagger X+2(L^\dagger L)\sqrt{1+\frac{m_1^2}{X^\dagger X}}\sqrt{1+\frac{m_2^2}{X^\dagger X}}\psik\\
W&=-\mu^2hX-\frac{m_1m_2}{h}\frac{L^2}{X}\point
\end{align}
We have reduced the model into one with two chiral superfields, and we can now minimize the potential. We find that at leading order, the potential is the same as in (\ref{eqn:runaway}):
\begin{gather}
\label{eqn:L(x)} L=\frac{\sqrt{\mu^2h^2-m_1m_2}}{\sqrt{m_1m_2}}X+O(\frac{m^3}{X^2})\psik\\
V=\frac{m_1m_2}{h^2}(2\mu^2h^2-m_1m_2)+\frac{m_1m_2}{4h^2X^4}(m_1^2-m_2^2)^2(\mu^2h^2-m_1m_2)+\dotsc \point
\end{gather}
Here $m$ stands for some masses in the Lagrangian.\\
The equations (\ref{eom_massivefields}) and (\ref{eqn:L(x)}) suggest a change of variables:
$R=\tilde R L,\:   P=\tilde P L,\:  L=\frac{1}{\sqrt 2}\tilde L$
In terms of these variables the K\"aher potential and superpotential for large $X$ are well approximated by
\begin{align}
K&=X^\dagger X+\tilde L^\dagger\tilde L+\tilde P^\dagger\tilde P+\tilde R^\dagger\tilde R\dotsc\psik\\
\label{eqn:suppot}W&=hX(\tilde R\tilde P-\mu^2)+\frac{m_1}{\sqrt 2}\tilde P\tilde L+\frac{m_2}{\sqrt 2}\tilde R\tilde L\point
\end{align}

The model has now a simple form which enables us to calculate the masses of the fields. We diagonalize the matrix of the quadratic terms of the Lagrangian to get the masses of the scalars and fermions. These are summarized in Table \ref{masstable}. The gauge field has mass squared of $2g^2\frac{(h^2\mu^2-m_1m_2)(m_1^2+m_2^2+2h^2X^2)}{m_1m_2}$.
The massless fermion is no other than the goldstino, arising from broken supersymmetry. The massless scalars are the pseudomoduli. Furthermore, there are two scalar fields that become very massive when $X$ gets large values and another two which become very light. We examine the behavior of the light fields in Appendix \ref{appendix:oneloop}.

\begin{table}[!h]
\centering
 \begin{tabularx}{\textwidth}{c|c}
\hline
\textbf{Scalars} & \textbf{Fermions} \\
 \hline \hline
 \\[0.1pt]
0  & \multirow{2}{*}{0}  \\
0 \\
\\[0.1pt]
\hline
\\[0.2pt]
$\frac{2\mu^2(2\mu^2h^2-m_1m_2)}{X^2}$  & \multirow{2}{*}{ $\frac{(2\mu^2h^2-m_1m_2)^2}{X^2}$} \\
$\frac{4h^4\mu^4-6h^2\mu^2m_1m_2+2m_1^2m_2^2}{h^2X^2}$ \\
\\[0.2pt]
\hline
\\[0.2pt]
 $h^2X^2+\frac{\mu^2h^2(m_1^2+m_2^2)}{m_1m_2}\pm2\mu^2h^2-\frac{1}{2}(m_1^2+m_2^2)$& \multirow{2}{*}{\footnotesize{$h^2X^2+h^2\mu^2\frac{m_1^2+m_2^2}{m_1m_2}-\frac{1}{2}(m_1^2+m_2^2)\pm(h^2\mu^2-m_1m_2)$}} \\
\\[1pt]
$h^2X^2+\frac{2h^2\mu^2(m_1^2+m_2^2)-m_1m_2(m_1^2+m_2^2\pm4(h^2\mu^2-m_1m_2))}{2m_1m_2}$\\
 \\[0.2pt]
 \hline
 \end{tabularx}
 \caption{Scalar and Fermion masses at the minimum (\ref{eqn:minima})}
\label{masstable}
\end{table}

\pagebreak
\section*{Acknowledgments}

We are grateful to L.~Carpenter, G.~Festuccia, Z.~Komargodski, and D.~Shih for useful discussions and collaboration at various stages of the project.   We are also grateful to the Institute for Advanced Study, NJ, USA, for hospitality during the course of this project. T.V  is supported by the ERC STG grant number 335182, by the Israel Science Foundation under grant number 884/11, by the United States-Israel Binational Science Foundation (BSF) under grant number 2010/629. In addition, the research of T.V is supported by the I-CORE Program of the Planning and Budgeting Committee and by the Israel Science Foundation under grant number 1937/12.  Any opinions, findings, and conclusions or recommendations expressed in this material are those of the authors and do not necessarily reflect the views of the funding agencies.

\appendix

\section{One loop corrections to the light fields}
\label{appendix:oneloop}

When calculating the fields masses along the runaway (Table \ref{masstable}) we got fields with $m\sim 1/X^2$. At large $X$, where the runaway is stabilized, these fields become light. We need to make sure that the minimum is not destabilized in this direction when considering the one loop corrections.

Since the massive fields are proportional to $X$, we can write the effective K\"ahler potential as
\begin{equation}
K_{eff}=Z_X(Q;\abs{X})X^\dagger X+Z_{\tilde P}(Q;\abs{X})\tilde P^\dagger \tilde P+Z_{\tilde R}(Q;\abs{X})\tilde R^\dagger \tilde R+\tilde L^\dagger \tilde L\psik
\end{equation}
where $Q$ is the RG scale and the $Z$'s are the wavefunction renormalizations.
Our minimum (\ref{eqn:minima}) is in the regime
\begin{equation}
m_0\ll\abs{X}\ll\Lambda\psik
\end{equation}
where $\Lambda$ is the cutoff scale of the low-energy theory. Therefore, we can estimate the effective potential in the following way \cite{Intriligator:2008fe}:
\begin{equation}
V_{eff}=Z_X(m_0;\abs{X})^{-1}\abs{F_X}^2+Z_{\tilde P}(m_0;\abs{X})^{-1}\abs{F_{\tilde P}}^2+Z_{\tilde R}(m_0;\abs{X})^{-1}\abs{F_{\tilde R}}^2+\abs{F_{\tilde L}}^2\psik
\end{equation}
which leads to:
\begin{equation}
V_{eff}=(1+2\gamma_X log(\frac{\abs X}{m_0}))\abs{F_X}^2+(1+2\gamma_{\tilde R} log(\frac{\abs X}{m_0}))\abs{F_{\tilde R}}^2+(1+2\gamma_{\tilde P} log(\frac{\abs X}{m_0}))\abs{F_{\tilde P}}^2+\abs{F_{\tilde L}}^2\point
\end{equation}
This gives rise to corrections of order one loop to the $\tilde L$ field's mass. Since the corrections only multiply the mass terms of the light fields, they cannot flip the sign of the potential, therefore these corrections will not change the minima that we found.

\bibliographystyle{utphys}
\bibliography{bib_summ}
\end{document}